\documentclass[a4paper,11pt]{article}
\pdfoutput=1

\usepackage{jheppub}

\usepackage{graphicx}
\usepackage{amsmath,amsfonts,amssymb}
\usepackage{color}
\usepackage{hyperref}
\usepackage{comment}
\usepackage{slashed}
\usepackage{tabularx}
\usepackage{enumitem}

\newcommand{\be}{\begin{equation}} 
\newcommand{\ee}{\end{equation}}
\newcommand{\bea}{\be\begin{aligned}} 
\newcommand{\eea}{\end{aligned}\ee}

\def\lsim{\mathrel{\raise.3ex\hbox{$<$\kern-.75em\lower1ex\hbox{$\sim$}}}}
\def\gsim{\mathrel{\raise.3ex\hbox{$>$\kern-.75em\lower1ex\hbox{$\sim$}}}}

\usepackage{array}
\newcolumntype{C}[1]{>{\centering\let\newline\\\arraybackslash\hspace{0pt}}m{#1}}

\newcommand{\tr}{{\rm tr}}

\newcommand{\td}{{\rm d}}
\newcommand{\pd}{\partial}

\makeatletter
\gdef\@fpheader{}
\g@addto@macro\bfseries{\boldmath}
\makeatother

\begin{document}

\title{Thermalization effects on the dynamics of growing vacuum bubbles}

\author[a]{Tomasz Krajewski,}
\emailAdd{tkrajewski@camk.edu.pl}
 \affiliation[a]{Nicolaus Copernicus Astronomical Center, ul. Bartycka 18, 00-716 Warsaw, Poland}
\author[b]{Marek Lewicki,}
\emailAdd{marek.lewicki@fuw.edu.pl}
\affiliation[b]{Faculty of Physics, University of Warsaw ul. Pasteura 5, 02-093 Warsaw, Poland}
\author[c,d]{Martin Vasar,}
\emailAdd{martin.vasar@ut.ee}
\affiliation[c]{Laboratory of High Energy and Computational Physics, NICPB, R{\"a}vala 10, 10143 Tallinn, Estonia}
\affiliation[d]{Institute of Physics, University of Tartu, W. Ostwaldi 1, 50411 Tartu, Estonia}
\author[c,e,f]{Ville Vaskonen,}
\emailAdd{ville.vaskonen@pd.infn.it}
\affiliation[e]{Dipartimento di Fisica e Astronomia, Universit\`a degli Studi di Padova, Via Marzolo 8, 35131 Padova, Italy}
\affiliation[f]{Istituto Nazionale di Fisica Nucleare, Sezione di Padova, Via Marzolo 8, 35131 Padova, Italy}\author[c]{Hardi Veerm\"ae}
\emailAdd{hardi.veermae@cern.ch}
\author[b]{and Mateusz Zych}
\emailAdd{mateusz.zych@fuw.edu.pl}

\abstract{We study the evolution of growing vacuum bubbles. The bubble walls interact with the surrounding fluid and may, consequently, reach a terminal velocity. If the mean free path of the particles in the fluid is much shorter than the bubble wall thickness, the fluid is locally in thermal equilibrium and the wall's terminal velocity can be determined by entropy conservation.
On the other hand, if local thermal equilibrium inside the wall cannot be maintained, the wall velocity can be estimated from the pressure impacted by ballistic particle dynamics at the wall. We find that the latter case leads to slightly slower bubble walls. Expectedly, we find the largest differences in the terminal velocity when the fluid is entirely ballistic. This observation indicates that the non-equilibrium effects inside walls are relevant. To study bubble evolution, we perform hydrodynamic lattice simulations in the case of local thermal equilibrium and $N$-body simulations in the ballistic case to investigate the dynamical effects during expansion. Both simulations show that even if a stationary solution exists in theory it may not be reached depending on the dynamics of the accelerating bubble walls.}

\maketitle

\section{Introduction}
\label{sec:intro}

Cosmological phase transitions are intriguing processes with rich phenomenological consequences. They can take place when two physically non-equivalent phases coexist, and the one occupying the Universe becomes energetically unfavourable. Phase transitions are usually modelled by dynamics of the order parameter, which in cosmological realisations is a scalar field that preserves the Lorentz symmetry. The phase transition is first order if the effective potential driving the evolution of the order parameter develops a barrier separating its minima corresponding to various phases at the moment when the hierarchy of the energy density of the phase which fills the Universe and the other one becomes inverted. The potential barrier forbids the field from smoothly evolving between the phases, and the transition has to proceed by quantum tunnelling~\cite{Coleman:1977py,Callan:1977pt} or thermal fluctuations~\cite{Linde:1980tt,Linde:1981zj}. Bubbles of the favourable phase (true vacuum) nucleate in the Universe filled with the other phase (false vacuum) as a result of these processes. They expand due to the release of latent heat which is given by the difference of the effective potential in two phases. If the order parameter couples to other constituents of the Universe, especially plasma of elementary particles, as it happens in many extensions of the Standard Model, the growth of bubbles can be significantly slowed down and even reach the constant subliminal velocity. Such terminal velocity is one of the most important parameters controlling the phenomenological implications of the phase transition. Determination of the value of the terminal velocity (and identification of the runaway scenario) is one of the unsolved problems in the theory of cosmological first-order phase transitions. In the current paper, we report the results of our efforts to compute such quantity from the first principles based on the content of the particle physics model standing behind the transition.

The first detection of gravitational waves (GWs) by LIGO and Virgo~\cite{LIGOScientific:2016aoc} opened new exciting possibilities for probing the Universe. While these observations continue~\cite{Abbott:2017vtc,Abbott:2017oio,Abbott:2017gyy,Abbott:2020khf,Abbott:2020tfl}, pulsar timing experiments have been reporting hints for a stochastic background~\cite{NANOGrav:2023gor,EPTA:2023fyk,Reardon:2023gzh,Xu:2023wog} at very low frequencies. The most likely culprit behind this signal are mergers of supermassive black hole binaries~\cite{NANOGrav:2023hfp, EPTA:2023gyr, Ellis:2023dgf}, but it is also possible that this background was generated in the early Universe~\cite{NANOGrav:2023hvm, EPTA:2023xxk, Ellis:2023oxs}, for example in a first-order phase transition. With new data flowing and optimistic prospects for the future due to number of planned experiments~\cite{Punturo:2010zz,Hild:2010id,Janssen:2014dka,Graham:2016plp,Audley:2017drz,Graham:2017pmn,LISA:2017pwj,Badurina:2019hst,Bertoldi:2019tck,Badurina:2021rgt,Ajith:2024mie}, hopes are rising for GWs as a probe of new physics.

The stochastic gravitational wave background generated during first-order phase transitions offers a unique glimpse into periods in the early Universe inaccessible through any other means~\cite{Caprini:2015zlo,Caprini:2019egz,LISACosmologyWorkingGroup:2022jok,Caprini:2024hue}. Expanding true vacuum bubbles play a central role in producing these signals. Ultrarelativistic walls generically produce a signal sourced by collisions of bubbles or relativistic fluid shells~\cite{Kosowsky:1992vn,Cutting:2018tjt,Ellis:2019oqb,Lewicki:2019gmv,Cutting:2020nla,Lewicki:2020jiv,Giese:2020znk,Ellis:2020nnr,Lewicki:2020azd,Lewicki:2022pdb}, while walls that reach the terminal velocity significantly below the speed of light allow fluid dynamics to play a more prominent role~\cite{Kamionkowski:1993fg,Hindmarsh:2015qta,Hindmarsh:2016lnk,Hindmarsh:2017gnf,Ellis:2018mja,Hindmarsh:2019phv,Ellis:2020awk,Jinno:2020eqg,Auclair:2022jod,Jinno:2022mie,Sharma:2023mao,RoperPol:2023dzg,Caprini:2024gyk}. Slower walls are also a necessity in typical models of electroweak baryogenesis~\cite{Kuzmin:1985mm,Cohen:1993nk,Rubakov:1996vz, Morrissey:2012db,Carena:2018cjh,Cline:2020jre,Cline:2021iff,Lewicki:2021pgr,Cline:2021dkf,Carena:2022qpf,Ellis:2022lft}. Therefore, gaining thorough understanding of the dynamics governing the evolution of bubbles and their connection to the physics of the surrounding matter is vital to fully utilize this observational window.

We employ a combination of semi-analytic methods supported by numerical simulations to investigate the dynamics of expanding true vacuum bubbles interacting with the surrounding plasma. On the one hand, we conduct hydrodynamic lattice simulations which assume local thermal equilibrium (LTE) and account for the changing particle mass across the bubble wall~\cite{Ignatius:1993qn, Kurki-Suonio:1995yaf, Kurki-Suonio:1996wfr, Hindmarsh:2013xza, Hindmarsh:2015qta}. The evolution of the order parameter field in lattice simulations (and at the same time, its coupling to the plasma) is driven by the thermally corrected effective potential. The LTE approximation requires the limit of zero mean free path of plasma particles, i.e. particles that scattered or interacted with the bubble wall instantaneously thermalize with the rest of the plasma. Under such an assumption, one can fully describe plasma as the perfect fluid using (spacetime-dependent) macroscopic parameters of temperature and velocity. On the other hand, we utilise particle-based simulations to capture nonequilibrium effects around the evolving phase interface~\cite{Lewicki:2022nba, Lewicki:2023mik}. In this way, it is possible to quantify the impact of nonequilibrium physics and to develop more effective techniques for handling deviations from thermal equilibrium in hydrodynamical simulations.

The paper is structured as follows: Sec.~\ref{sec:theory} reviews the theoretical background and outlines the setup for numerical approaches. In Sec.~\ref{sec:velocity}, we will consider three different approaches for determining the wall velocity depending on degree of thermalization in the system. Results are presented in Sec.~\ref{sec:results}, and we conclude in Sec.~\ref{sec:concl}.
We use units in which $\hbar = c = 1$ and the $+---$ metric signature.

\section{Theoretical considerations}
\label{sec:theory}

\subsection{Fluid dynamics}
\label{sec:fluid_dynamics}

We consider a model consisting of a potentially non-thermal fluid\footnote{As we do not assume thermal equilibrium or the absence of counter-streaming, it would be more accurate to talk about a bath of particles. As the non-thermal nature can be understood from the context, will not make this distinction in terminology.} coupled to a real scalar $\phi$ responsible for the phase transition. The fluid consists of particles that are massless in the false vacuum and acquire a mass in the true vacuum. In thermal equilibrium, this is fully captured by the thermal potential and the equation of state (EOS) of the fluid.

The stress-energy tensor of the bubble--fluid system can be divided into the stress-energy tensor of the scalar $\phi$ and the fluid, $T^{\mu\nu} = T_{\phi}^{\mu\nu} + T_{f}^{\mu\nu}$. Due to the interactions between the scalar and the fluid, the split of $T^{\mu\nu}$ is not unique. We define $T_{\phi}^{\mu\nu}$ as the free scalar contribution,
\be\label{eq:stress_energy_scalar}
    T_{\phi}^{\mu\nu} 
    \equiv \pd^{\mu} \phi \pd^{\nu} \phi + g^{\mu\nu}\left[-\frac{1}{2} \pd_\alpha \phi \pd^\alpha \phi + V_{0}(\phi) \right] \,,
\ee
where $V_0$ represents the scalar potential in absence of the fluid. The stress-energy tensor for a fluid in which particles follow a distribution $f_i({\vec p})$ is
\be\label{eq:stress_energy_fluid}
    T_f^{\mu\nu}  = \sum_i \int \frac{\td^3 k}{(2\pi)^3} \frac{k^{\mu} k^{\nu}}{E} f_i({\vec k})\, .
\ee
The dependence on $\phi$ arises implicitly through the field dependence of the masses of the particles, $E \equiv E(\vec k, m_i(\phi))$.

\begin{figure}
    \centering
    \includegraphics[width=0.8\textwidth]{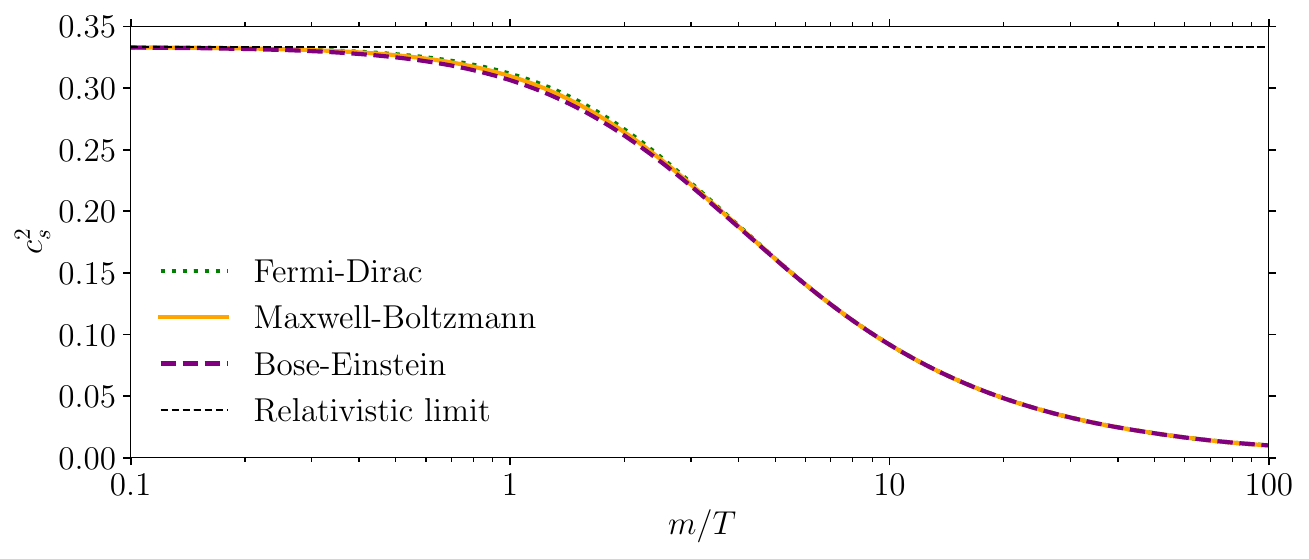}
    \caption{The dependence of the speed of sound $c_s^2$ on the mass-temperature ratio for three thermal distributions: Fermi--Dirac, Maxwell--Boltzmann and Bose--Einstein.}
    \label{fig:LTE_EoS}
\end{figure}

For a perfect fluid in local thermal equilibrium, we have
\be
    T_f^{\mu\nu} = w \, u^{\mu}u^{\nu} - p \, g^{\mu\nu} \,,
\ee
where $u^{\mu}$ is the 4-velocity of the fluid (normalized such that $u_\mu u^\mu = 1$), $w$ is the enthalpy density and $p$ is the pressure. All quantities can be derived from the temperature dependence of pressure as
\be\label{def:w,s}
    w = \rho + p = T s, \qquad
    s = \frac{\partial p}{\partial T} \,,
\ee
where $\rho$ is the energy density and $s$ is the entropy density. In the fluid's rest frame, the equilibrium distribution of species $i$ is $f_{i} = g_i /(e^{E/T} - \sigma_i)$, where $g_i$ is the number of degrees of freedom and $\sigma_i=1$ for bosons and $\sigma_i=-1$ for fermions. This gives 
\be\label{eq:p_phi}
    p \equiv \sum_i \int \frac{\td^3 k}{(2\pi)^3} \frac{k^2}{3E}f_i(E) = -\sum_i \frac{g_i T^4}{2\pi^2}J_{\sigma_i}\left(\frac{m_i(\phi)}{T}\right)\, ,
\ee
where the thermal functions are given by
\be
    J_{\sigma}(x) 
    = 
    - \sigma^{-1} \int_{0}^{\infty} \!\!\td y \,y^2\ln\left( 1 - \sigma e^{ - \sqrt{y^2 + x^2}} \right) \,.
\ee
The Maxwell-Boltzmann case is obtained in the limit $\sigma \to 0$. For simplicity, in both lattice and $N$-body simulations, we will consider the EOS corresponding to the Maxwell-Boltzmann distribution, that is, classical statistics. This is justified as the speed of sound $c_s^2 \equiv \partial p/\partial \rho$, shown in Fig.~\ref{fig:LTE_EoS} for the Fermi--Dirac, Maxwell--Boltzmann and Bose--Einstein distributions, depends very mildly on the underlying statistics of the particles. The effect of the statistics is quantified explicitly in Sec.~\ref{sec:full_LTE} in the context of stationary states.

In general, the scalar field obeys the equation of motion (EOM)~\cite{Dolan:1973qd, Moore:1995si, Lewicki:2022nba}
\bea\label{eq:eom_FVB+PP}
    \Box \phi + V_0' 
    +  \sum_i \partial_\phi m_i^2(\phi) \int \frac{\td^3 k}{(2\pi)^3} \frac{f_i(\vec k)}{2E} 
    = 0\, .
\eea
Given that there is a frame (the rest frame of the fluid) in which $f$ depends only on the energy of the particles, $f = f(E)$, we obtain
\bea
    \int \frac{\td^3 k}{(2\pi)^3} \frac{f_i(E)}{2E} = -\pd_{m_{i}^2} p \,.
\eea
Consequently, the scalar field evolution is determined by the thermal potential,
\be \label{eq:eom_field}
    \Box \phi + V' = 0, \qquad 
    V(\phi,T) \equiv V_0(\phi)  - p(\phi, T).
\ee
As this property holds whenever $f=f(E)$ in the fluid's rest frame, this assumption is weaker than requiring thermal equilibrium.

To obtain equations of motion for the fluid, we apply conservation of the energy-momentum tensor of the system, $\nabla_\mu T^{\mu \nu} = 0$ which implies $\nabla_\mu T^{\mu \nu}_{\phi} = - \nabla_\mu T^{\mu \nu}_{f}$. Together with the EOM of the scalar field~\eqref{eq:eom_FVB+PP}, we then obtain 
\be \label{eq:eom_munu_1}
    \nabla_\mu T^{\mu \nu}_{\phi} 
    = -\sum_i \partial^\nu m_i^2(\phi) \int \frac{\td^3 k}{(2\pi)^3} \frac{f_i}{2E} 
\ee
and
\be \label{eq:eom_munu_2}
    \nabla_\mu T^{\mu \nu}_{f} = -\pd^{\nu}\phi \,  \pd_{\phi} p \,.
\ee
In conclusion, assuming spherical symmetry, $u^\mu = (\gamma, \gamma v, 0, 0)$ the evolution of the perfect fluid is governed by
\bea \label{eq:eom_fluid}
    \pd_t (w \gamma^2) + \frac{1}{r^2}\pd_r(r^2 w \gamma^2 v) - \pd_t p  
    &= \pd_t \phi \pd_\phi p \,, \\
    \pd_t (w \gamma^2 v) + \frac{1}{r^2}\pd_r(r^2 w \gamma^2 v^2) + \pd_r p 
    &= -\pd_r \phi \pd_\phi p \,.
\eea
The hydrodynamical approach comprises solving Eqs.~\eqref{eq:eom_field} and \eqref{eq:eom_fluid} numerically, starting from an initial bubble profile. We describe details of the numerical computation in Appendix~\ref{app:hydro}.

\subsection{Scalar sector}

In the hydrodynamic lattice simulations, we consider the general quadratic potential, $V_0(\phi) = -\mu^2\phi^2/2 - \mu_3 \phi^3/3 + \lambda\phi^{4}/4$ with $\lambda > 0$, that has minima at $\phi = 0$ and $\phi = v$. By trading $\mu$ with $v$ and defining $\eta = \mu_3/(\lambda v) - 1$, so that the local maximum is at $\phi = \eta v$, the potential can be expressed as
\be \label{eq:V0} 
    V_0(\phi) = \lambda \left( \frac{\eta}{2}v^2\phi^2 - \frac{1+\eta}{3}v\phi^3 + \frac{1}{4}\phi^{4}\right) \,.
\ee
For $\eta < 1/2$, the true vacuum lies at $\phi = v$, while for $\eta > 1/2$, it lies at the origin, and for $\eta = 1/2$, the minima are degenerate. We consider $\eta < 1/2$. 

The potential is constructed so that $V_0(0) = 0$. The potential energy difference between the vacua,
\be
    \Delta V_0 = \frac{\lambda v^4}{12}(1 - 2\eta) \,,
\ee
has a mild dependence on $\eta$ when $\eta \lesssim 1/2$, while the height of the barrier, $V_{0}(\eta v) = (\lambda v^4/12) (2-\eta)\eta^3$, diminishes rapidly as $\eta \to 0$. As the barrier is generated at the tree level, we expect that including quantum corrections (the Coleman--Weinberg potential) will not significantly affect our results, and thus, we will omit them in this study. 

The mass of the scalar in the false vacuum is $m_0^2 = \lambda v^2 \eta$ whereas in the true vacuum it is $m_v^2 = \lambda v^2 (1-\eta) > m_0^2$. As an illustrative example, we include an additional fermionic field with a field-dependent mass 
\be
    m_{\psi}^2 = y^2 \phi^2 \,.
\ee
We consider the situation where the field is initially at $\phi = 0$ and the transition to the true vacuum occurs at the temperature $T \ll m_0$ so that the fluid at the transition temperature contains only massless fermions.

\subsection{Stationary bubbles}
\label{sec:stationary}

In this study, we are mostly interested in scenarios in which the bubbles will reach a stationary state with a terminal wall velocity $v_w < 1$. In such stationary scenarios, the dimensionality of the fluid equations \eqref{eq:eom_fluid} can be generally reduced. For the fluid profiles in the stationary state, the full numerical solution of Eqs.~\eqref{eq:eom_field} and \eqref{eq:eom_fluid} can be easily approximated using matching equations along the bubble wall~\cite{Espinosa:2010hh,Giese:2020znk}. In this approach, the wall velocity is typically assumed to be one of input parameters. However, it is possible to determine the wall velocity by considering an additional condition that determines the pressure on the wall.  

Two approximations can be made: First, the field outside the wall will occupy a minimum of the potential and can be assumed to be constant, i.e., $\pd_\mu \phi = 0$. Second, as the fluid equation~\eqref{eq:eom_fluid}  does not contain any explicit scales, the stationary solution can be assumed to depend only on $\xi \equiv r/t$. With these approximations, Eqs.~\eqref{eq:eom_fluid} can be arranged into
\bea \label{eq:statstate}
    \pd_\xi v &= \frac{2v}{\xi} \frac{\gamma^{-2}}{1-\xi v}\left(\frac{\tilde{v}^2}{c_s^2(w)} - 1\right)^{\!-1} , \\
    \frac{\pd_\xi w}{w} &= \left( \frac{1}{c_s^2(w)} + 1 \right) \tilde{v} \gamma^2 \pd_\xi v \,,
\eea
where $\tilde{v} \equiv (\xi-v)/(1-\xi v)$. These equations give two fixed points: if $\xi=c_s$, then $v=0$ and if $\xi=1$ then $v=1$. However, the fluid velocity must approach zero at some point after the wall. This can only happen with some discontinuous jump. Three possibilities exist: detonations, deflagrations and hybrids (supersonic deflagrations). Notably, such a discontinuity is not a simplifying approximation and is expected from the asymptotic nature of the solution. In particular, the proper physical width of the scalar profile $\Delta r$ must remain constant in the stationary state. Thus, the corresponding size in the $\xi$ coordinate will vanish as the bubble approaches a stationary state, that is, $\Delta \xi \equiv \Delta r/t \to 0$ when $t\to \infty$.

In addition, to join two fluid profiles inside and outside the wall, we can use conservation of the energy-momentum of the system, Eqs.~\eqref{eq:stress_energy_scalar} and \eqref{eq:stress_energy_fluid}. This gives rise to the matching equations 
\bea\label{eq:matching_equations}
    w_{+} \bar{v}_{+}^2 \bar{\gamma}_{+}^2 + p_{+} - \Delta V_0 
    &= w_{-} \bar{v}_{-}^2 \bar{\gamma}_{-}^2 + p_{-}  \ , \\
    w_{+} \bar{v}_{+}\bar{\gamma}_{+}^2 
    &=  w_{-} \bar{v}_{-}\bar{\gamma}_{-}^2,
\eea
where $\Delta V_0 = V(0, T) - V(\phi_0, T)$. These conditions hold in the frame of the bubble wall (respective variables are denoted with $\bar{x}$). To close the system of equations, an EOS must be defined. For simplicity, to account for the changing speed of sound in the true vacuum, we assume Maxwell--Boltzmann distribution, $f\propto e^{-E/T}$.  

In addition, one can use a simple EOS - an extension of the bag model~\cite{Giese:2020znk} (further on ``extended bag model''), where fluid is assumed to be ultrarelativistic and the speed of sound in each phase is constant. Thermodynamic quantities are defined as 
\begin{gather}
    \begin{aligned}
    p_+ = \frac{1}{3}aT^4 - \Delta V \,,\qquad\qquad &&\rho_+&  = aT^4 + \Delta V \,,\\
    p_- = \frac{1}{3} b T^{\nu} \,,\ \ \ \qquad\qquad\qquad&&\rho_-&  = \frac{1}{3}b(\nu-1)T^\nu\,,
    \end{aligned}
\end{gather}
where $+$ and $-$ show if a~quantity is defined outside or inside the bubble respectively and $a$, $b$ are some constants with $a > b$. The parameter $\nu$ sets the speed of sound in the phase inside the bubble with
\be
    \nu \equiv 1 + 1/c_s^2\,.
\ee
We assume that the speed of sound outside the bubble is close to $c_s^2=1/3$. More about this approach can be found in~\cite{Giese:2020znk}, from which we took the code snippet to compare the results.

Usually, thermodynamic quantities are known only far away from the wall outside the vacuum bubble. Thus, we know only boundary conditions to the fluid Eqs.~\eqref{eq:statstate}. To find the full solution, we will thus adapt the shooting method when necessary. Three qualitatively different scenarios must be considered:

\begin{enumerate}
    \item \textbf{Detonations} possess only a rarefaction wave trailing the bubble wall. They are computationally the simplest as the variables in front of the wall must match the asymptotic ones: $v_{\xi=1} = v_{+} = 0$ and $w_{\xi = 1} = w_{+}$; and $v_{-}$ and $w_{-}$ are easily found from the matching equations. Detonations are possible only in the regime where $v_w > c_{s, +}$.
    
    \item \textbf{Deflagrations} possess only a shock wave in front of the wall. As the fluid behind the wall is at rest, we have $v_{-} = 0$. In this scenario, the shooting method is required to adapt boundary conditions at $\xi=1$. The discontinuity of the shock wave is fixed by $\tilde{v}(\xi_{sh}, v_{sh})\xi_{sh} = c^2(\xi_sh)$. This corresponds to limit $\td v/\td \xi \rightarrow -\infty$. Deflagrations exist when $c_{s, -} < v_w$. 

    \item \textbf{Hybrids} have a distinctive rarefaction wave behind and a shock wave in front of the wall, thus combining features of both cases discussed above. Numerical and analytical studies have shown that stable solutions exist when $\tilde{v}_{-} =c_{s, -}$~\cite{Kurki-Suonio:1995yaf, Megevand:2014yua}. 
\end{enumerate}

\section{Determining the wall velocity}
\label{sec:velocity}

Let us now consider additional conditions that would allow us to fix the speed of the wall. We will consider three different scenarios with varying degree of non-thermal behaviour:
\begin{enumerate}
    \item LTE in the entire system,
    \item LTE only away from the wall, but ballistic motion inside the wall,
    \item fully ballistic fluid (free-streaming particles).
\end{enumerate}
The ballistic scenarios are realized when the mean free path of the particles is either longer than the thickness of the wall (the scenario 2) or the thickness of the plasma shell (the scenario 3). The full ballistic approach represents the extreme case in which the particle-particle interactions in the fluid are essentially negligible and the particles do not thermalize during the phase transition. The partly ballistic scenario corresponds to the intermediate case, where the fluid particles' motion is affected only by changes in their mass as they cross the phase interface. We will not consider potential non-thermal friction arising from higher order particle-wall interactions, such as soft particle emission~\cite{Bodeker:2017cim,Azatov:2020ufh,Gouttenoire:2021kjv,Hoche:2020ysm,BarrosoMancha:2020fay}, which tends to be more relevant for ultrarelativistic walls, which we do not consider here.

\subsection{Local thermal equilibrium across the wall}
\label{sec:full_LTE}

Entropy is conserved if LTE is maintained across the bubble wall~\cite{Hindmarsh:2020hop, Ai:2021kak}. This follows directly from the equations of motion. The gradient of the stress-energy tensor of a perfect fluid~\eqref{eq:stress_energy_scalar} can be rewritten as
\bea
    u_{\nu}\nabla_\mu T^{\mu \nu}_{f} 
    &= T \nabla_\mu (s u^{\mu}) 
    + u^{\mu} \nabla_\mu T \overbrace{\left(w/T - \partial_T p\right)}^{=0} - u^{\mu} \nabla_\mu \phi \partial_\phi p\,,
\eea
where we used Eq.~\eqref{def:w,s} $s = w/T = \partial_T p$ and $u_{\nu}\nabla_\mu u^{\nu} = 0$, $u_{\mu}u^{\mu} = 1$. On the other hand, Eq.~\eqref{eq:eom_munu_2} gives $u_{\nu}\nabla_\mu T^{\mu \nu}_{\phi} = u^{\mu}\nabla_\mu \phi \partial_\phi p$. Thus, combing this with the conservation of the total energy-momentum $\nabla_\mu (T^{\mu \nu}_{f} + T^{\mu \nu}_{\phi}) = 0$, we obtain
\be\label{eq:S_current=0}
    \partial_\mu (s u^{\mu}) = 0\,.
\ee
In the wall frame, this implies the additional matching condition 
\be\label{eq:S=const}
    s_- \gamma_- v_- = s_+ \gamma_+ v_+\,,
\ee
which allows us to determine the bubble's wall velocity for a given model. In the hydrodynamical simulations, we confirmed that the conservation of entropy current \eqref{eq:S_current=0} holds within expected numerical errors.

\subsection{Ballistic dynamics at the wall}

Scenarios with ballistic dynamics at the wall can be studied using the thin wall approximation. Crucially, when the wall thickness is much smaller than the size of the bubble, momentum exchange between the wall and a particle can be derived purely from energy-momentum conservation and does not depend on the wall profile. This validates the use of the thin-wall approach for such scenarios.

The radius $R$ of the thin-wall bubble obeys the EOM~\cite{Ellis:2019oqb,Lewicki:2022nba}
\be\label{eq:eom_R}
    \ddot{R} + 2 \frac{1-\dot{R}^2}{R} = \frac{\big(1-\dot{R}^2\big)^{3/2}}{\sigma} \left(\Delta V_0 - \Delta P\right) \ , 
\ee
where $\Delta V_0$ is the potential energy difference between two vacua omitting thermal corrections and $\sigma$ is the bubble wall tension. The thermal corrections are accounted for by the pressure difference $\Delta P$ across the bubble wall caused by the energy transfer between the particles and the wall. Omitting particle self-interactions inside the wall and the soft-particle emission, the pressure difference $\Delta P$ is given by~\cite{Lewicki:2022nba}
\be\label{eq:DP}
    \Delta P 
    \!=\!\! \int \frac{\td^3 p}{(2\pi)^3} \sum_{j \in \pm 1}  f_{j}(p)\frac{(n{\cdot} p)^2}{E_i} \theta(-j n{\cdot} p) \left[ 
    \mathcal{T}_j(n{\cdot} p) \left(1\!-\!\sqrt{1\!-\!j \frac{\Delta m^2}{(n\!\cdot\!p)^2}}\right) 
    \!+\! 2 \mathcal{R}_j(n{\cdot} p) \right],
\ee
where $j=-1$ and $j=+1$ refer, respectively, to regions right behind and right in front of the wall, $f_{j}(p)$ denotes the distribution of particles' momenta in these regions, $\Delta m^2 \equiv m_{-}^2 - m_{+}^2 > 0$ is the change in the particle mass when it enters the bubble from outside, $\mathcal{R}$ and $\mathcal{T} = 1 - \mathcal{R}$ are reflection and transmission coefficients and $n^\mu = (v_w \gamma_w,\gamma_w,0,0)$ is the bubble wall normal vector. 

The particles crossing the wall are affected only by the change in mass $m(\phi)$ due to the change of the scalar field value. Due to quantum effects, particles energetically allowed to cross the wall may still be reflected with probability $\mathcal{R}(n\cdot p)$.

In our ballistic simulations, we will assume (see Appendix~\ref{app:reflection_and_transmission})
\be\label{eq:R_coeff}
    \mathcal{R}_{-}(u) = \theta(\Delta m^2 - u^2)\,, \qquad
    \mathcal{R}_{+} = 0 \, ,
\ee
that is, we omit reflections when the wall crossing is kinematically allowed. We expect that this choice will not significantly affect our conclusions.

By~\eqref{eq:eom_R}, the terminal wall velocity $v_w$ is determined by 
\be\label{cond:v_w}
    \Delta P = \Delta V_0 \,.
\ee
In the scenario where we assume LTE away from the wall, but not inside it, we solve the fluid profiles in the stationary state from~\eqref{eq:statstate}. This gives the distribution of particles $f_{j}(p)$ colliding with the wall, allowing us to evaluate~\eqref{eq:DP}. Finally, the terminal wall velocity can be obtained by iterating \eqref{eq:statstate}, \eqref{eq:DP} and \eqref{cond:v_w}.

\subsection{Fully ballistic fluid}

The terminal velocity for effectively non-interacting fluids can be estimated analytically using the ballistic limit because the particles scattered from the wall do not heat the particles about to scatter from the wall. The fluid's density and velocity profiles in front are far from equilibrium. They can be separated into two components that do not interact with each other: a nearly thermal component at rest that has not yet interacted with the wall and an out-of-equilibrium component moving away from the expanding bubble. The pressure arises from the first component, and its properties match the properties of the fluid far from the bubble, which makes it possible to estimate $\Delta P$ and thus the terminal velocity via \eqref{cond:v_w}.

Neglecting quantum corrections to the reflection coefficient, i.e., imposing Eq.~\eqref{eq:R_coeff}, and assuming that the momenta follow a Maxwell-Boltzmann distribution without fixing the number density, the pressure is given by~\cite{Lewicki:2023mik}
\bea\label{eq:DP_appr}
    \Delta P(T,v_w) 
    = \frac{\rho}{3} \frac{(1+v_w)^2}{1-v_w} \left[1 - G_{\Delta P}\left(\frac{\Delta m}{T} \sqrt{\frac{1-v_w}{1+v_w}}\right)\right] \,,
\eea
where $v_w$ is the wall velocity and $G_{\Delta P}\left(x\right) \equiv (1/4)\left(e^{-x}(2+2x+x^2) + x^2 K_2(x)\right)$ with $K_2$ being the modified Bessel function. From Eqs.~\eqref{cond:v_w} and \eqref{eq:DP_appr}, we see explicitly that the terminal velocity is determined by two ratios: $\Delta V_0/ \rho$ and $\Delta m/T$. In the limit of relativistic walls, $v_w \to 1$, Eq.~\eqref{eq:DP_appr} recovers the well-known result of Ref.~\cite{Bodeker:2009qy}.

\section{Results}
\label{sec:results}

\subsection{Profiles}
\label{sec:plasma_profiles}

\begin{figure}
    \centering
    \includegraphics[width=\columnwidth]{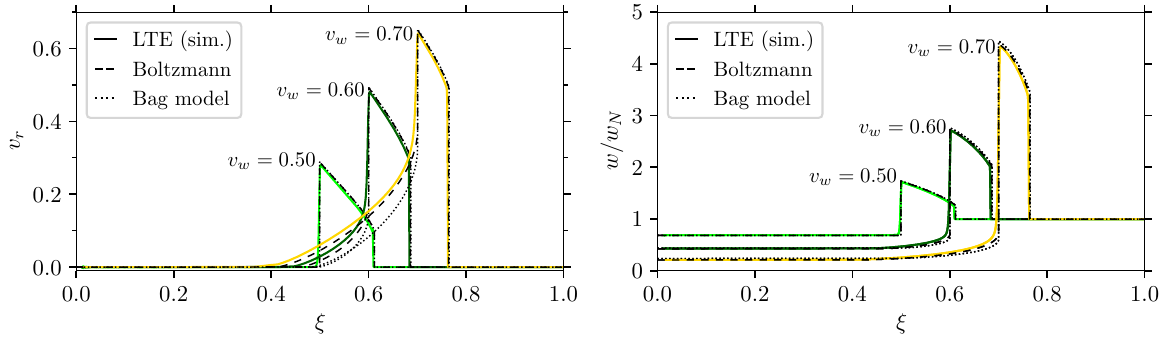}
    \includegraphics[width=\columnwidth]{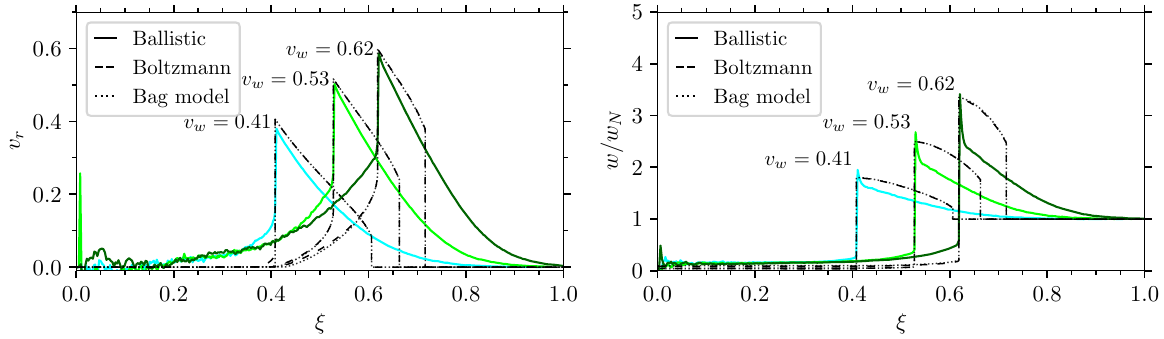}
    \caption{Fluid velocity and enthalpy profiles from hydrodynamic simulations (solid lines in the upper panel) and ballistic simulations (solid lines in the lower panels) compared with the stationary state approach for the Maxwell--Boltzmann EOS (dashed lines) and the bag EOS (dotted lines). Colours correspond to parameters indicated by the stars in Fig.~\ref{fig:main}.}
    \label{fig:profiles}
\end{figure}

We compare fluid profiles from three different methods. Firstly, we simulate bubble walls with hydrodynamical simulations assuming LTE everywhere. Secondly, we calculate the stationary states using both the bag and the Maxwell--Boltzmann EOS, taking the wall velocity as an input parameter. Thirdly, we perform $N$-body simulations, including particles interactions. A detailed description of the $N$-body simulation can be found in the Appendix~\ref{app:ballistic}.

We show the velocity and enthalpy profiles obtained with these three methods in Fig~\ref{fig:profiles}. The upper panels show that profiles obtained with the hydrodynamic simulations match very well with the stationary state method using the Maxwell--Boltzmann EOS. This is expected as the hydrodynamical simulations must asymptotically approach the profiles predicted by the stationary state method. This is consistent with earlier studies based on the bag equation of state~\cite{Krajewski:2023clt, Krajewski:2024gma}. Noticeable differences appear in the rarefaction wave profiles between the bag and the Boltzmann EOS in the stationary state approach. These differences can be fully attributed to the fact that the speed of sound inside the bubble is temperature-dependent for the Boltzmann EOS but not for the bag model. The region outside the bubble is populated by radiation ($p = \rho/3$), and all profiles agree well with each other.

In the lower panel of Fig.~\ref{fig:profiles}, the results of $N$-body simulations are compared with profiles obtained from (hydrodynamic) stationary state methods for the same velocities of bubble walls as obtained in the simulations. The significant differences are clearly visible. In particular, the stationary state method predicts the formation of shocks where the plasma shell exceeds the speed of sound. A finite mean free path of particles is considered in the $N$-body simulations, and therefore, the shock front does not form.

\subsection{Bubble-wall velocity in the interacting and non-interacting limits}

Let us turn our attention to the terminal wall velocity in the three scenarios outlined in Sec.~\ref{sec:velocity}. We will describe the system with two parameters: the phase transition strength 
\be
    \alpha = \frac{\Delta V_0}{\rho(T)}\, ,
\ee
and the ratio $\Delta m / T$, where $T$ denotes the fluid temperature far outside the bubble.

\begin{figure}
    \centering
\includegraphics[width=\textwidth]{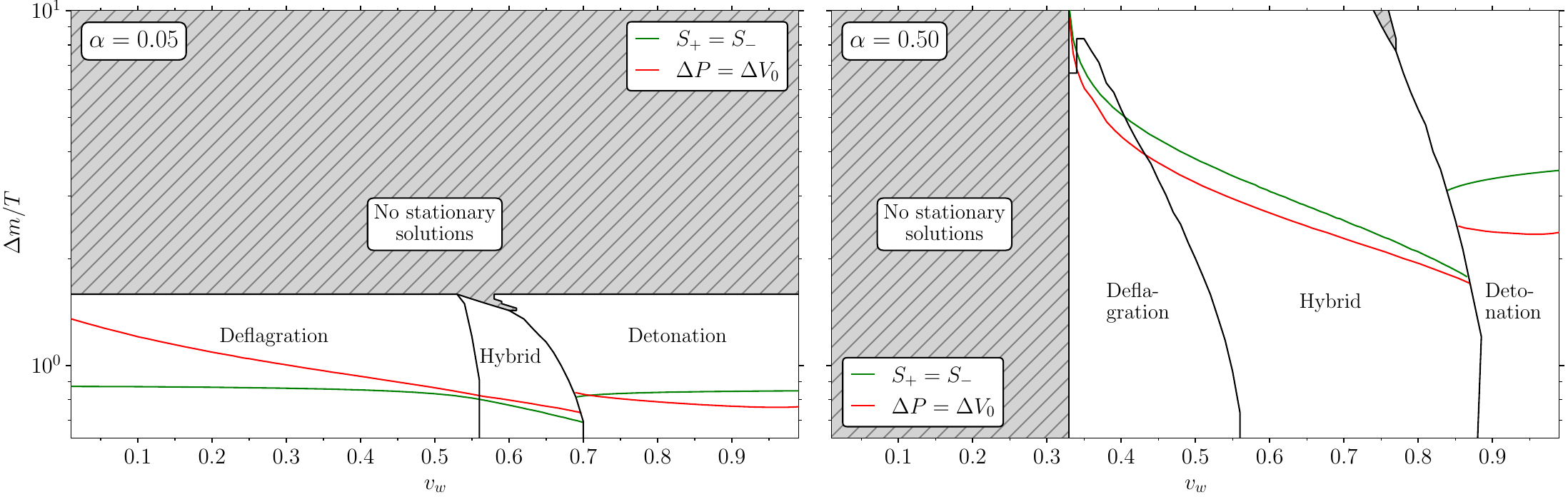}
    \caption{Solution types when solving Eq.~\eqref{eq:statstate} with matching equations Eq.~\eqref{eq:matching_equations} as constraints. One additional constraint is needed to fix the solution: entropy conservation (green line) or $\Delta P=\Delta V_0$ (red line). The hatched region shows the parameter space where a stationary state solution was not found.}
    \label{fig:LTE_scan}
\end{figure}

To map the landscape of stationary states, we scanned over these two parameters and the wall velocity $v_w$. We show two slices of the parameter space corresponding to different $\alpha$ in Fig.~\ref{fig:LTE_scan}. For each set of parameters, we determined the type of the profile and calculated fluid variables behind and in front of the wall. This way, we can easily apply an extra constraint, assuming it only depends on $v_{\pm}$ and $T_{\pm}$, that determines the wall velocity. The green and red curves in Fig.~\ref{fig:LTE_scan} show the wall velocity obtained respectively with entropy conservation, corresponding to the assumption of LTE in the entire system, and with the condition $\Delta P = \Delta V_0$, corresponding to the scenario of having LTE only away from the wall, but ballistic motion inside the wall.

\begin{figure}
    \centering\includegraphics[width=1\linewidth]{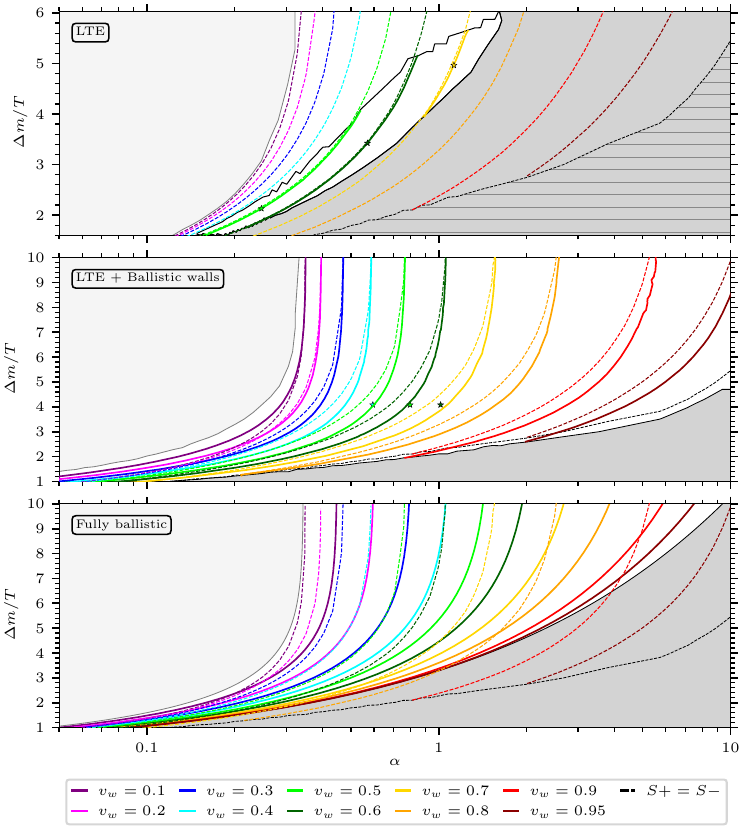}
    \caption{Results of three scenarios: LTE based on entropy conservation (top panel), LTE away from the wall and ballistic dynamics inside it (middle panel), and feebly interacting particles (bottom panel). The light and dark grey regions show collapse and runaway regions, respectively, for each scenario. In the top panel, the dark region shows runaways according to the hydrodynamical simulations. The black dashed line shows the runaway region calculated using the stationary state method and assuming entropy conservation (LTE everywhere). The coloured stars show points corresponding to the profiles shown in Fig.~\ref{fig:profiles}.}
    \label{fig:main}
\end{figure}

\begin{figure}
    \centering
    \includegraphics[width=\textwidth]{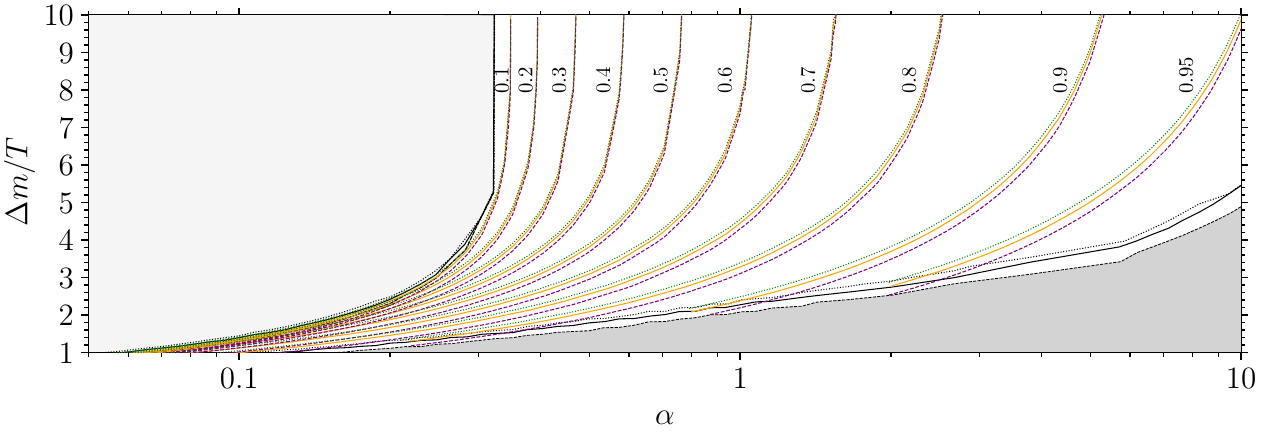}
    \caption{Wall velocity dependence on the phase transition strength $\alpha$ and $\Delta m/T$ for three thermal distributions (continuous - Maxwell--Boltzmann, dashed - Bose--Einstein, dotted - Fermi--Dirac). Stationary solutions can be found between the black lines. The dark grey region contains runaways, and in the light grey region the bubble collapses for Bose distributions.}
    \label{fig:LTE_EoS_2}
\end{figure}

In Fig.~\ref{fig:main}, we compare the wall velocities in different scenarios as a function of $\alpha$ and $\Delta m / T$. We chose only deflagrations and hybrids for the stationary state method because stationary detonations were not found in the hydrodynamical simulations. Although the stationary state method finds stationary detonations, they are degenerate with solutions describing deflagrations or hybrids (see Fig.~\ref{fig:LTE_scan}).

As seen from the top panel of Fig.~\ref{fig:main}, the results of the hydrodynamical simulations and the stationary state method assuming entropy conservation Eq.~\eqref{eq:S=const}, agree very well in all cases for which a stationary state was found using both approaches.
The stationary state method predicts runaway behaviour in the hatched region in the top panel of Fig.~\ref{fig:main}, while the hydrodynamical simulations show that certain stationary states satisfying matching conditions cannot be realized as a terminal state of nucleated bubbles.  We discuss the existence of the stationary states further in Sec.~\ref{sec:runway_solutions}. In Fig.~\ref{fig:LTE_EoS_2}, we show the terminal velocity obtained with the stationary state method in LTE for different thermal distributions. We see that the dependence on the type of thermal distribution is mild.

In the middle panel of Fig.~\ref{fig:main}, we see that the scenario with ballistic dynamics at the wall and LTE away from it results in slightly lower velocities than the fully LTE scenario. This is because the violation of LTE inside the wall is associated with entropy production. The two scenarios converge to the same asymptote in the non-relativistic limit ($\Delta m > T$). This asymptote corresponds to a scenario where $v_{+}$ approaches $v_w$, meaning that the fluid appears to be not moving in the wall frame.

\begin{figure}
\centering\includegraphics[width=1.0\columnwidth]{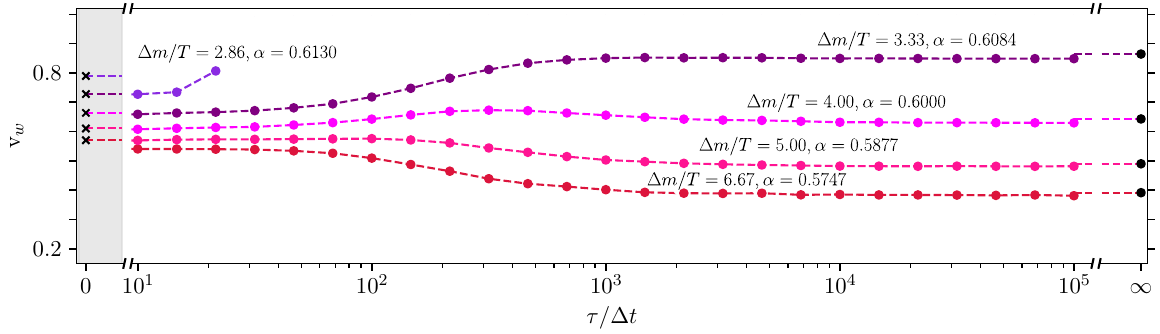}
    \caption{The effect of thermalization on the terminal wall velocity for different $\Delta m / T$ and $\alpha$. The  strength of particles' interactions is characterized by the timescale $\tau$, shown in units of the simulation timestep $\Delta t$. The limits $\tau \to 0$ and $\tau \to \infty$ correspond to instant thermalization and free particles, respectively.}
    \label{fig:Nbody_thermalization}
\end{figure}

The bottom panel of Fig.~\ref{fig:main} shows that in the scenario with feebly interacting particles, the wall velocity can be very different. To see how thermalization time qualitatively changes the stationary wall velocity, we performed $N$-body simulations with different mean free times. The results are shown in Fig.~\ref{fig:Nbody_thermalization} for five choices of $\Delta m/T$. Results with large $\tau$ (long mean free path) correspond to feebly interacting matter, and the numerical results agree well with the corresponding analytical solution, Eq.~\eqref{eq:DP_appr}, shown as the $\tau\to\infty$ limit. For small $\tau$, the system asymptotically approaches the LTE limit. The resulting $\tau\to 0$ limit in Fig.~\ref{fig:Nbody_thermalization} is obtained with the stationary state method using the condition $\Delta P = \Delta V_0$ to solve for the wall velocity. We see that the $N$-body simulations predict slightly lower wall velocities at short mean free paths than what is given by the limit of LTE outside the wall. Even though the low mean free time results do not agree well with the analytics, they still imply that thermalization time has an important effect on stationary wall velocity. This disagreement is likely a result of nonequilibrium effects. We also see that in some cases, even the existence of a stationary solution depends on whether thermalization is fast enough.

\subsection{Can all stationary solutions be reached?}
\label{sec:runway_solutions}

\begin{figure}
    \centering    
    \includegraphics[width=1\columnwidth]{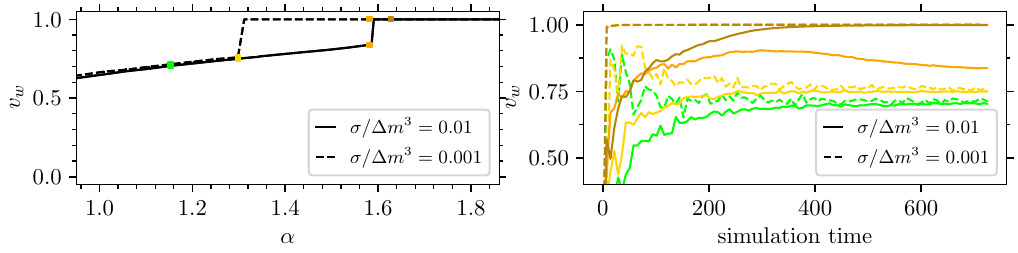}
    \caption{Dependence of the terminal wall velocity on the surface tension term $\sigma$ for $\Delta m/T = 3.4$. \emph{Left panel:} terminal velocity dependence on phase transition strength parameter $\alpha$ for two different $\sigma/\Delta m ^3$. \emph{Right panel:} the evolution of the wall velocity evolution corresponding to points depicted on the left plot.}
    \label{fig:surface_tension_max_velocity}
\end{figure}

As can be seen on the top panel of Fig.~\ref{fig:main}, not all the stationary states found using the LTE matching condition~\eqref{eq:S=const} are realized in hydrodynamical simulations. This phenomenon was already described in~\cite{Krajewski:2024gma} but for the model characterized by a different EOS. It was observed, and we confirmed this scenario, that even for reasonable initial conditions of recently nucleated critical bubbles in a homogeneous plasma, the bubble wall accelerates beyond the Jouguet velocity before the heated fluid shell in front of the bubble, characteristic for deflagrations and hybrids, is fully formed. In such a scenario, the hydrodynamical obstruction does not work and the evolution leads to a runaway scenario, even though a stationary state exists. Such behaviour was observed for strong enough transitions and is depicted in Fig.~\ref{fig:main} with the dark grey shaded region without hatching. This region contains stationary states found via the LTE matching conditions shown by the coloured dashed lines. According to the LTE matching conditions (and also numerical hydrodynamical simulations), no stationary states exist in the hatched region.

A similar phenomenon was observed in the ballistic simulations. From the EOM~\eqref{eq:eom_R} it can be seen that the wall tension parameter $\sigma$ does not change the terminal velocity when $R\rightarrow \infty$ and $\Delta P$ stays the same. With the particle-particle interactions turned on, the fluid thermalizes, making $\Delta P$ time-dependent. To check this scenario, we ran $N$-body simulations by varying $\alpha$ ($\Delta V_{0}$) and $\sigma$ with fixed $\Delta m/T$. We confirmed, as is shown in the left panel of Fig.~\ref{fig:surface_tension_max_velocity}, that changing $\sigma$ did not alter the terminal wall velocity, but affected whether the stationary solution exists. The right panel of Fig.~\ref{fig:surface_tension_max_velocity} shows that in the $N$-body simulations, the $\sigma$ value influences how fast the wall velocity changes. The wall accelerates faster for smaller $\sigma$.

\section{Conclusions}
\label{sec:concl}

We have addressed the key question in the physics of cosmological first-order phase transitions: whether a stationary state of expansion bubbles exists and what is the terminal velocity of their walls when this state is reached. This determines the phenomenological predictions of the transition since faster walls emit stronger gravitational waves, while slower walls are expected to produce higher matter-antimatter asymmetry in scenarios of electroweak baryogenesis. The terminal velocity is much harder to predict than other parameters of the transition. Even though, a significant effort has been undertaken by the community to calculate it from first principles, the final conclusion was still not clear.

Concretely, we have studied the thermalisation effects on bubble wall dynamics. Understanding its role is crucial for determining the bubble wall dynamics. We have considered three qualitatively different cases, depending on the mean free path of particles in the fluid:
\begin{enumerate}
    \item mean free path much shorter than the bubble wall thickness---the fluid is in local thermal equilibrium (LTE) everywhere,
    \item mean free path longer than the bubble wall thickness, but much shorter than the thickness of the fluid shells around the wall---the fluid is in LTE everywhere except inside the wall,
    \item mean free path longer than the thickness of the fluid shells---the fluid is not in thermal equilibrium.
\end{enumerate}
In the first two cases, the stationary fluid profiles around the wall can be approximated from hydrodynamic equations with matching conditions that ensure energy-momentum conservation. The terminal velocity is then determined in the first case from entropy conservation and in the second case by matching the ballistic pressure with the potential energy difference that drives the bubble expansion. In the third case, the fluid around the wall is far from thermal equilibrium, the ballistic pressure arises only from those particles that have not yet interacted with the bubble and the terminal velocity is obtained by matching the pressure with the potential energy difference. 

To study whether a terminal velocity is reached, we have also performed hydrodynamic LTE and $N$-body ballistic simulations. Our LTE simulations and the matching method agree very well on the predicted wall velocity whenever the simulation predicts a final steady state. The main difference is that the simulations predict a runaway rather than a steady state for the strongest transitions. In addition, we have shown that, in the LTE regime, Fermi--Dirac statistics predict slightly slower and Bose--Einstein statistics faster wall speeds than the Maxwell--Boltzmann distribution. 

The intermediate scenario in which LTE is violated only inside the bubble wall predicts slightly smaller wall velocities. This is expected as the violation of LTE inside the wall is associated with the production of entropy which can only slow the wall down. Our results show that the LTE approximation may not be sufficient to predict the terminal velocity. 

The fully ballistic scenario, corresponding to the case of feebly interacting particles, predicts significantly different wall velocities from the other two cases. Interestingly, we have found that the increase in the mean free path of particles can either slow the wall down or increase its terminal velocity. For strong transitions where this approximation is most relevant, the walls are typically faster for non-interacting particles.

We have found very good agreement in profiles of plasma shells between our hydrodynamical perfect fluid simulations and the matching method whenever the former converges to a stationary state rather than a runaway. The ballistic simulations with a finite mean free path predict, as expected, significantly different profiles without a sharp shock front. However, the hydrodynamical parameters at the peak are similar between these cases when comparing walls with the same velocity.

Finally, both hydrodynamic and ballistic dynamical simulations showed that the evolution of the bubble needs to be tracked to predict its final fate. This is because a stationary state, even if it exists, is not necessarily realized. We have found that some initially accelerating bubbles can reach a high velocity before a stationary heated plasma shell can be formed. We observed that this can lead to runaways even if a stationary configuration exists in theory. Numerical simulations of evolving bubbles with more accurate modelling of the nonequilibrium behaviour of the plasma need to be performed in the future to pinpoint the nature of a stationary state of expansion of bubbles.

\vspace{20pt}
\noindent
\emph{Note added} -- As our manuscript was finalised, Ref.~\cite{Ai:2024uyw} dealing with very similar topics appeared.

\section*{Acknowledgments} 

This work was supported by the Estonian Research Council grants PRG803, PSG869, RVTT3 and RVTT7 and the Center of Excellence program TK202. The work of M.L. and M.Z. was supported by the Polish National Agency for Academic Exchange within the Polish Returns Programme under agreement PPN/PPO/2020/1/00013/U/00001.
The work of V.V. was partially supported by the European Union's Horizon Europe research and innovation program under the Marie Sk\l{}odowska-Curie grant agreement No. 101065736.
The work of T.K. was supported by the Polish National Science Center grant 2019/33/B/ST9/01564.

\appendix

\section{Scalars in a bubble background}
\label{app:reflection_and_transmission}

Neglecting the expansion of the Universe, the evolution of the bubble in vacuum follows from the action
\be\label{eq:Sphi}
    S_{\phi} = \int \td^4 x \left[ \frac{1}{2} \pd_\alpha \phi \pd^\alpha \phi - V_0(\phi) \right]\, ,
\ee
where $V_0$ denotes the potential without thermal corrections. Consider now the scalar quanta (particles) on a bubble background around the phase interface. Splitting the field into a classical background and quantum fluctuations $\phi = \phi_0 + \hat{\phi}$, the free scalar particles obey
\be
    S^{(0)}_{\hat{\phi}} 
    = \int \td^4 x \left[ \frac{1}{2} \pd_\alpha \hat{\phi} \pd^\alpha \hat{\phi} - \frac{1}{2} m_{\rm eff}^2(\phi_0(x)) \hat{\phi}^2 \right]\,.
\ee
These particles will contribute to the thermal surroundings. Although effective mass is negative at the local maximum, i.e., $m_{\rm eff}^2(\phi \eta) = - \lambda v^2 \eta (1-\eta) \leq 0$, there is no tachyonic instability inside the wall, as we will show. The quantized field is given by
\be
    \hat{\phi} = \int \frac{\td^3 k}{(2\pi)^3} \, \frac{1}{\sqrt{2E_{\vec k}}} \hat a_{\vec k} \, \phi_{\vec k}({\vec x},t) + h.c. \, ,
\ee
where the mode functions are determined from $\Box \phi_{\vec p} + m_{\rm eff}^2 \phi_{\vec p} = 0$.\footnote{Mode functions are normalized as $\int \td^3x  \phi^*_{\vec p}\phi_{\vec q} = (2\pi)^3\delta^{(3)}(\vec p - \vec q)$.} Although the momenta are not conserved close to the wall, the modes can be labelled in terms of the asymptotic momenta in the true or false vacua.

It is simplest to work in the wall frame. Given that the thickness of the wall is much smaller than the radius of the bubble, we can further neglect the curvature of the wall, that is, assume a planar wall. For definiteness we will set up the problem with the wall positioned at $x=0$, with the 4-momentum and mass at $x\to -\infty$ given by $\vec k = (k, {\vec k}_\parallel)$ and $m_-$ and at $x\to \infty$ by $\vec q = (q, \vec k_\parallel)$ and $m_+$, with $m_- < m_+$ and ${\vec k}_\parallel$ denoting the conserved momenta parallel to the wall. Wall frame energy conservation relates the remaining momentum components 
\be\label{eq:E_k_wall}
    E^2 = m_-^2 + k^2 + {\vec k}_\parallel^2 = m_+^2 + q^2 + {\vec k}_\parallel^2\,.
\ee
The mode equation is separable so that we can define
$
    \phi_{\vec k} = e^{-i E_{\vec k} t + i \vec k_\parallel \vec x_\parallel} \phi_k(x)\,,
$
which leaves us with the one-dimensional problem
\be\label{eq:phik}
    -\partial_x^2 \phi_{k} + (m_{\rm eff}^2(x) - m_-^2 - k^2) \phi_{k} = 0 \,
\ee
for the mode function in the direction perpendicular to the wall.

To proceed, we will approximate the wall profile as
\be\label{eq:phi0}
    \phi_0(x) = \frac{v}{2} \left(1+\tanh\left(\frac{x}{2w}\right)\right),  \quad
    w = \frac{1}{\sqrt{2 \lambda}v}\,.
\ee
The wall frame width $w$ is found by minimizing the action \eqref{eq:Sphi} with the above ansatz with respect to $w$. It is independent of $\eta$. Moreover, the ansatz~\eqref{eq:phi0} solves the EOM of~\eqref{eq:Sphi} exactly when the potential is degenerate ($\eta=1/2$). In the background~\eqref{eq:phi0}, the mode equation~\eqref{eq:phik} admits two independent analytic solutions $\tilde \phi_{\pm k}(x)$ with
\bea
    \tilde\phi_{k}(x) 
    = &\frac{e^{ikx}}{(1+e^{x/w})^2}{}_{2}F_{1}(-2 + i w (k+q), -2 + i w (k-q), 1 + 2 i w k, -e^{x/w})\,,
\eea
where ${}_{2}F_{1}$ denotes the hypergeometric function. These solutions satisfy $\phi_k(x \to -\infty) \propto e^{ikx}$, $\tilde\phi^*_{k}(x) = \tilde\phi_{-k}(x)$ and are symmetric under $q \to -q$. The spectrum consists of two types of modes:
\begin{itemize}[leftmargin=*]
    \item When $|k| > \Delta m \equiv \sqrt{m_+^2 - m_-^2}$, then $q$ is real and the particles can cross the barrier. The mode functions are 
    \be 
        \phi_{\pm k} \propto \tilde \phi_{\pm k}\, .
    \ee

    \item When $|k| \leq \Delta m$, then $q$ is imaginary, and the particles cannot cross the barrier. The mode function is determined by demanding that the growing mode $e^{x|q|}$ vanishes when $x \to \infty$ and, at $x \to -\infty$, the mode consists of both the incident and reflected waves contributing equally, that is,
    \be
        \phi_{k} = (e^{i \theta_{k,q}} \tilde\phi_{k} + e^{-i \theta_{k,q}} \tilde\phi_{-k})/\sqrt{2}\, ,
    \ee
    where $\theta_{k,q}$ is a momentum dependent phase. This mode function is invariant under $k \to -k$.
\end{itemize}
The solution for the mode equation implies that the reflection coefficient is
\be\label{eq:reflection}
    \mathcal{R} 
    = \left|
    \frac{\sinh(\pi w (k - q))}{\sinh(\pi w (k + q))}
    \right|^2 \,
\ee
for particles crossing the phase boundary from either direction. Besides the possibility of reflections when $|k| \geq \Delta m$, the scalar quanta can be modelled as classical point particles. For wider walls, $w \Delta m \gtrsim 1$, the reflection coefficient is exponentially suppressed when $|k| \gg \Delta m$, so that reflections can be safely neglected in the $k \geq \Delta m$ region. For thinner walls, $\Delta w \ll 1$, $\mathcal{R} \approx (k-q)^2/(k+q)^2$ and the reflection coefficient decreases polynomially with $k$. However, even in the $w\to 0$ limit, the reflection coefficient is $\mathcal{R} = 1/2$ already at $k = 1.015\Delta m$, that is, reflections dominate at most in a narrow momentum shell $k/\Delta m \in [1,1.015]$. They are relevant only for particles barely capable of crossing the phase boundary.

Finally, a remark about maintaining LTE around the wall is on order. The thermal density matrix $\hat \rho = e^{-\beta \hat H^{(1)}}$ is determined from the (leading order) Hamiltonian of scalar quanta
\be
    \hat H^{(1)} = \int \frac{\td^3 k}{(2\pi)^3} E_{\vec k} \hat a^{\dagger}_{\vec k} \hat a_{\vec k} \, ,
\ee
where ${\vec k}$ labels states.\footnote{Here, we only consider scattering states from the continuous part of the spectrum and neglect potential bound states that would be localized inside the wall as such states would mostly be related to the evolution of the wall itself.} Since states crossing the wall are inherently not localized, the momentum labels will be specified in the vacuum with asymptotic mass $m_-$, as in \eqref{eq:E_k_wall}. The logarithm of the partition function $Z = \tr \hat \rho$ takes the usual form
\be\label{eq:lnZ_phi}
    \ln Z 
    = - V \int \frac{\td^3 k}{(2\pi)^3} \ln(1-\exp(-\beta E_k))\,.
\ee
In thermal equilibrium, away from the wall, the "+" side does contain states with  $|k| \geq \Delta m$ since the $|k| < \Delta m$ are exponentially damped in the "+" vacuum, and we obtain the well-known quantum distributions. Importantly, even if the effective mass squared can be negative inside the wall, $m_{\rm eff}^2(x) < 0$, the energy of the particles always satisfies $E \geq \min(m_{-},m_{+})$. Additionally, due to modes whose wavelength exceeds the wall thickness, it is not always meaningful to impose \emph{local} equilibrium within the wall, especially if such "non-local" modes dominate the distribution. Nevertheless, we will not attempt to account for these quantum effects in this work and assume the thermal potential of the form~\eqref{eq:p_phi}.

\section{Ballistic simulations with self-interacting particles}
\label{app:ballistic}

We simulate ballistic scenarios using $N$-body simulations in 3D space. $N$ particles are generated in the false vacuum with Boltzmann distribution given by temperature $T_n$, which corresponds to temperature far away from the bubble wall. In addition, we initialize the bubble with an initial radius $R_0$. Given $N$, $R_0$ and $T_n$,    we can fix the simulation box size, with the edge length $L_{\text{sim}}$, such that energy density in the false vacuum corresponds to Boltzmann distribution energy density. 

We used the multi-particle collision dynamics (MPC) method to simulate the thermalization process between the particles. After each timestep, particles are assigned to a collision cell with a length $R_{\text{cell}}$. We limited the maximum number of particles in a cell to 2. If more than two particles would appear in a cell, we created a new cell to be filled up and carried out this process until all particles were assigned a cell. Then, the particles inside the same cell would interact. The reason for fixing the particle count to 2 is that we could not make the algorithm reach the Maxwell-Boltzmann distribution with a higher number of particles in a cell. For each cell, we determine the probability of collision happening in the cell. The collision is performed as a random rotation of the momentum vector around a random axis in the cell's zero-momentum frame. This method leaves total energy unchanged. MPC method was used as it is parallelizable, which makes simulating collisions faster. 

A collision in each collision cell must also occur with a probability proportional to $\sqrt{E1 E_2}$~\cite{Lewicki:2022nba}. In addition, another probability $\propto \exp(\Delta t / \tau)$  is used, where $\tau$ is the thermalization time parameter and $\Delta t$ is timestep. Furthermore, we introduced additional density-dependent probability necessary to simulate shock wave-like behaviour. The latter was hard to achieve as collision cell resolution was not good enough, but increasing the number of cells was not possible due to GPU memory limitations.  Parameter $\tau$ is proportional to thermalization time and limits $\tau \rightarrow 0$ and $\tau \rightarrow \infty$ correspond to instant thermalization and no interaction, respectively. Even if $\tau =0$, the $N$-body simulation algorithm does not result in LTE in front of the bubble, as collisions in the algorithm happen once per timestep, thus generating some numerical out-of-equilibrium effects around the bubble wall.

The accuracy of the collision simulation is limited by two parameters: temporal resolution, $\Delta t$, and spatial resolution, $R_{\text{cell}}$. Changing either comes with a computational and memory cost. Changing timestep scales simulation approximately linearly in simulation runtime. Changing cell length scales simulation cubically as there are $N_{\text{c}}N_{\text{cell}}^3$ collision cells in the simulation, where $N_{\text{c}}$ is how many collision cells are placed at the same position and $N_{\text{cell}}$ is the number of cells per dimension. In this work we used $\Delta t = R_b / 10^4$ and $R_{\text{cell}} = 2R_b / 91$ where $R_b$ is simulation's boundary radius and $N_{\text{cell}} = 91$.

Each timestep consists of three subsequent procedures:
\begin{enumerate}
    \item Each particle's location is evolved using the Euler method, and collisions with the bubble wall are resolved. For particles leaving the simulation box, we use periodic boundary conditions $x \rightarrow x \pm L_{\text{sim}}$, with the sign depending on from which side the particle tries to leave the simulation box.
    \item Apply the collision algorithm.
    \item Bubble wall evolution according to Eq.~\eqref{eq:eom_R}.
\end{enumerate}

Simulation runs till the bubble wall reaches the simulation boundary or some unphysical condition (or error) is met.

A drawback of the current implementation is the inefficient scaling of the collision cell count. This could be fixed by using a collision cell only if it is not empty. This means that the maximum number of cells would be only $N/2$ where $N$ is particle count. Currently the minimum cell count is $(2R_{\text{sim}}/R_{\text{cell}})^3$. Also, implementing the collision algorithm with any number of particles in a cell could alleviate the problem. We also found that this algorithm violates causality due to the finite size of the collision size, which makes the instantaneous collisions effectively non-local. This causes a measurable effect on only very fast bubble walls with a speed of $v_w \gtrsim 0.7$ above which the results of ballistic simulations of self-interacting particles become less reliable.

\section{Set-up of hydrodynamical simulations}
\label{app:hydro}

For the numerical evolution, we use the variables $Z\equiv w\gamma^2v$ and $\tau\equiv w\gamma^2 - p$. We discretize and numerically solve the equations~\eqref{eq:eom_field} and~\eqref{eq:eom_fluid}. We use the finite element method, both in time and space. To discretise in space, we used the discontinuous Galerkin method. Our elements are just intervals of length $\Delta r$ in the computational domain $[0, R]$. We used the value of $R$ large enough to guarantee that the wall of the bubble is far from $r=R$ during the whole simulation, thus the choice of the boundary condition at this point does not influence the results.

Since fluxes in Eq. \eqref{eq:eom_fluid} are determined in terms of both conserved and so-called primitive variables $v$, $T$ (and derived from them $p$) one has to determine primitive ones from $\phi$, $\tau$ and $Z$ which are evolved in the code. To do so, we combine equations of state to find
 \be
     \tau + p(\phi, T) - \frac{1}{2}\left(w(\phi, T) +\sqrt{w(\phi,T)^2+ 4Z^2} \right) = 0\,,
 \ee
which we solve using the Raphson-Newton method for the value of the temperature $T$. Then, $w$ and $p$ can be directly computed, and the velocity $v$ can be computed simply by inverting the definition of $Z$. For a detailed description of the setup, see~\cite{Krajewski:2023clt, Krajewski:2024gma}.

\bibliographystyle{JHEP}
\bibliography{main}

\end{document}